\documentclass[10pt]{article}
\usepackage[utf8]{inputenc}
\usepackage{amsmath,amsfonts,amssymb,latexsym}
\usepackage[T1]{fontenc}
\newtheorem{satz}{Theorem}[section]

\newtheorem{assumption}[satz]{Assumption}
\newtheorem{defi}[satz]{Definition}

\newtheorem{bem}[satz]{Remark}
\newtheorem{lemma}[satz]{Lemma}
\newtheorem{koro}[satz]{Corollary}
\newtheorem{conclusion}[satz]{Conclusion}
\newtheorem{ob}[satz]{Observation}

\newtheorem{propo}[satz]{Proposition}

\newcommand{\mcal}{\mathcal}

\newcommand{\tit}{\textit}

\newcommand{\beq}{\begin{equation}}
\newcommand{\eeq}{\end{equation}}

\begin{document}
\thispagestyle{empty}
\begin{center}
\vspace*{1.0cm}
{\Large{\bf  About the Incompleteness of the Semiclassical\\ Picture in the  BH-Information Paradox\\ and the Quantum Hair emerging from the\\Coupling of Quantum Matter and Quantum Space-Time Degrees of Freedom}}
\vskip 1.5cm
 
{\large{\bf Manfred Requardt}}

\vskip 0.5cm

Institut fuer Theoretische Physik\\
Universitaet Goettingen\\
Friedrich-Hund-Platz 1\\
37077 Goettingen \quad Germany\\
(E-mail: requardt@theorie.physik.uni-goettingen.de or muw.requardt@googlemail.com) 

\end{center}
\begin{abstract}
We argue that the semiclassical analysis of the black hole information paradox is incomplete and has to be completed by an explicit entanglement of matter and quantum gravity degrees of freedom. We study in  detail the evaporation process from beginning to end in the light of our extension and show that a pure initial state remains pure over the full evaporation process, including the final state which remains after the black hole has completely evaporated. By the same token we show how the quantum hair is encoded in the micro structure of the (final) gravitational state which is entangled with the (final) radiation state.
\end{abstract}
\newpage
\section{Introduction}
About 40 years ago Stephen Hawking argued that the usual rules of quantum mechanics do not seem to hold in a process in which a black hole (BH) forms and then completely evaporates (\cite{Hawking1},\cite{Hawking2}). According to Hawking the paradox is the following. His \tit{semiclassical} calculation shows that the emitted radiation is exactly thermal. This does not present a problem as long as there do exist enough states inside the BH since \tit{entanglement} between internal and external quantum states is quite natural. But in the end, when the BH has completely evaporated, what remains is a thermal radiation state with nothing left to couple to. That is, an initial \tit{pure} state has evolved into a \tit{mixed} thermal state. This cannot happen in quantum theory!   
Since then this issue has been intensely debated with quite a few resolutions being proposed by some authors which then were, on the other hand, frequently criticized by other scientists but a universally accepted explanation does not have emerged in our view (see however the so-called \tit{BH-war}, \cite{BHWar}). 
 
We recently came upon some papers which claim to have solved the BH-information paradox, \cite{Hsu1}, \cite{Hsu2}, by coupling explicitly matter and gravitational  degrees of freedom (DoF).While the applied techniques and procedures are different, the invoked big picture is similar to our paper \cite{Requ_BH-info}, i.e., in contrast to the ordinary scenario of the semiclassical picture it is assumed that there is built up an explicit entanglement between radiation DoF and gravitational DoF and which survives the evaporation of the BH, so that the final state is an entangled pure state of radiation modes and gravity modes. In a wider context may also belong papers by Kay, \cite{Kay1}, \cite{Kay2}, while they emphasize slightly different aspects of the coupling between quantum matter DoF and gravitational DoF. Furthermore, in e.g. \cite{Donelly} a quantum field is introduced, which is explicitly dressed by gravitational DoF, representing, in our view, some sort of effective field theory compared to a perhaps more fundamental quantum gravity reality.

As we do not intend to give a review of this quite facetted (battle) field of the BH-information paradox with papers going in the hundreds we will only cite a few representative papers mainly from the ``classical'' period. A quite up to date review is \cite{Harlow}. The respective material is also represented for example in \cite{BH},\cite{Preskill},\cite{Page},\cite{Carlip}or \cite{Huijse}, just to mention a few sources. 

The approach of Hawking may be called \tit{semiclassical}, implying that one essentially studies quantum field theory in a (slowly changing) classical background. We will not discuss the careful but somewhat tedious original treatment by Hawking employing conditions of the quantum field on minus or plus null-infinity together with some boundary conditions on the event horizon of the BH. This is an approach which remains completely in the space-time outside the BH horizon. On the other hand one may be interested in a treatment which deals also with the interior of the BH, as it is done for example in the so-called \tit{nice slices} approach using slices which cover both the exterior and the interior region. We are convinced (as are quite a few other colleagues) that it would be of considerable help if one finds an appropriate treatment of what is going on in the interior of the BH during the evaporation process.
However one may entertain the philosophy that it should be possible to analyze the problems in each appropriate system of reference provided one uses the correct and adapted notions, see for example the remarks in \cite{Wein}.

As far as we can see, most of the researchers in this field (at least as far as the high energy physics side is concerned) are of the opinion that a \tit{pure infalling state} remains pure in the evaporation process, that is, the full evolution is unitary. On the other hand, the typical suggested solutions appear to be frequently quite contrived and sometimes seem to invoke  what should rather be derived from clear premises. A typical argument is that somehow early and late Hawking radiation have to be strongly entangled because the overall radiation state has to be pure so that ``information'' can finally be recovered from the presumed subtle correlations. Given the process by which the Hawking pairs are created and the huge space-time distance between early and late Hawking radiation  we are very sceptical whether sufficient correlations of this type do really exist.

The nice slice argument and a possible \tit{non-local} behavior is given a closer inspection in e.g. \cite{Susskind1},\cite{Lowe1},\cite{Lowe2},\cite{Giddings1},\cite{Giddings2},\cite{Giddings3},\cite{Mathur1}. It was however observed in \cite{Marolf1},\cite{Marolf2},\cite{Unruh1} that some of the strong entanglement assumptions are in fact too strong, more precisely, the assumption that both early and late Hawking radiation and exterior and interior modes of the late Hawking radiation are strongly entangled, and violate a rigorous result of quantum information theory, i.e. the so-called \tit{entanglement monogamy}. Put differently, in the view of the authors the most conservative resolution of the paradox is the existence of a \tit{firewall} for an infalling observer, thus violating the cherished \tit{equivalence principle} of general relativity. 

The main reason, however, why most researchers stick to the opinion that the evaporation process remains unitary over the full course of time is the so-called \tit{ADS-CFT correspondence}. As the bulk theory can be mapped onto a boundary field theory which is unitary by definition, the same must hold for the bulk theory. However, this argument from \tit{duality} is quite indirect (as the correspondence is relatively subtle concerning the details of the mapping) and does not really explain how the bulk physics has to be modified. A recent and quite detailed analysis of these points has been given in \cite{Papadodimas1} and \cite{Papadodimas2}. 

In various of the above cited papers it is speculated or even argued that some sort of non-locality will play a certain role in this tangle of different and competing explanations. While many researchers are aware of the possibility that the semiclassical picture may represent a problem it turns out to be quite difficult to really pinpoint the place where this philosophy may break down. In our view, however, the analysis is incomplete right from the beginning which we will explain in the following section.

We will show that non-locality inevitably does occur if one sticks to a pure particle picture, as is usually done in this BH-context. In contrast to that we will argue in favor of a mixed particle-gravitational mode picture which is essentially local and where the gravitational modes replace the infalling particle.
\begin{bem} What concerns quantum phenomena we adopt in this paper a \tit{substantival} point of view (as to this notion cf. e.g. Sklar in \cite{Sklar}). I.e., we assume for example that vacuum fluctuations are really permanently present in the vacuum even if not being detected by an observer.
\end{bem}
%%%%%%%%%%%%%%%%%%%%%%%%%%
\section{The Zero Energy Universe as a Result of the Coupling of Quantum Matter and Quantum Space-Time}
In this section we want to provide arguments for the strict coupling and correlation between \tit{quantum matter} (QM) excitations in the quantum vacuum and corresponding \tit{hole excitations} in the \tit{quantum space-time} (QST) substratum. It is our aim to argue that in contrast to the semiclassical philosophy this coupling extends to the field of BH-physics and remains crucial in this weak field scenario.

However, as we plan to treat this wider and more fundamental context in greater detail elsewhere, we will only provide, in broad strokes, some cursory arguments and remarks in order to better understand the relevance of this more primordial picture for the assumed weak field scenario around the event horizon of a BH.

In a first step we will introduce the concept of the \tit{zero energy universe}. Our central assumption is that before all these phase transitions which are introduced in the various more or less fundamental theories there was one really primordial phase transition (PT),  the \tit{mother of all phase transitions} so to speak, as a result of which both \tit{(quantum) space time} (QST) and \tit{(quantum) matter} (QM) did emerge from some irregular and wildly fluctuating substratum, QX, which does not support any stable structures and has no \tit{near} or \tit{far order}. It is assumed to consist of a large network of elementary degrees of freedom which essentially interact with each other in a very erratic form while ST denotes the smooth classical space-time manifold and M classical matter.

For some reason, which we do not specify in this paper, a structural phase transition or large fluctuation does occur in the network, in the consequence of which an extended submanifold QST of more coherently interacting DoF begins to emerge providing some smoothness and some degree of near/far order. This embedded submanifold we will associate with our (quantum) space time QST. In e.g. \cite{Requardt_GRG} or \cite{Requardt_Lumps} we developed a more rigorous picture of such a network scenario plus geometric phase transition.

In this process, which is a phase transition of first kind, some degree of order is established which yields a lowering of entropy and production of \tit{latent heat}. this latent heat transforms a certain fraction of the DoF of the original QX which are not involved in the formation of the system QST into what we experience as quantum matter QM.

That is, to put it briefly, we have a primordial phase transition
\beq QX\stackrel{PT}{\to}(Q)ST/(Q)M \eeq
in which the \tit{internal energy} of QST is lowered compared to the original QX by an amount $\triangle E$ while this same amount is transferred to QM. That is
\beq E(QX)= E(QST)+E(QM) \eeq

While this brief description of the presumed primordial processes may seem quite speculative, one should note that the ideas, underlying the \tit{inflationary} picture of the \tit{Big Bang} are very similar (cf. e.g. \cite{Guth1} and \cite{Guth2}). We want in particular stress the similarity to the \tit{zero-energy universe} idea (see \cite{Rosen},\cite{Tryon},\cite{Vilenkin} or the calculation in\cite{Wheeler} concerning the gravitational energy of a pair of vacuum fluctuations).

In a next step we are going to corroborate this picture by discussing a, at first glance, quite unrelated point which occurs frequently in the discussion of vacuum fluctuations. In a somewhat heuristic picture it is often argued that (pairs of) virtual particles or mere excitations pop out of the quantum vacuum, but as they have a life time which falls below the energy-time uncertainty relation, cannot be detected by an observer. What is in our view sometimes overlooked is the fact that the vacuum is usually considered in quantum field theory to be the ground state of the theory, that is, it is an eigenstate, $\Omega$. But eigenstates of the Hamiltonian have necessarily zero standard deviation, i.e.
\beq <(H-<H>_{\Omega})\Omega|(H-<H>_{\Omega})\Omega>=0  \eeq
or with $H\Omega=0$:
\beq <H^2>_{\Omega}=0  \eeq
which is a stronger condition than $<H>_{\Omega}=0$.
\begin{bem} We discussed this problem of vacuum fluctuations already in \cite{Requardt_Wormhole} sect.2.
\end{bem}

Hence, according to the  standard interpretation of quantum theory combined with spectral theory, this implies that in each individual observation process we find that the total sum of positive and negative energy fluctuations in the vacuum must exactly vanish and not only in the mean. That is, at each moment the global pattern of energy fluctuations in the quantum vacuum is an array of highly correlated positive and negative elementary energy fluctuations which add up to zero.
\begin{conclusion}If we assume that the various elementary fluctuations do occur on a short microscopic scale in space with their sum adding up to zero energy this appears as 
a fine tuned behavior.
\end{conclusion}
\begin{bem} It should be mentioned that Hawking in \cite{Hawking1} invoked exactly this picture of particle pairs near the event horizon with the (virtual) particle, having negative mass, falling into the singularity while the other having positive mass escaping to infinity.
\end{bem}

In contrast to such a particle oriented picture we favor the idea that only a combination of QM and QST yields the complete picture. In other words, we introduce instead of the incomplete QM vacuum an extended vacuum state consisting of an empty QM section comprising the positive matter energy excitations and a QST or gravitational section of energy levels which are assumed to be mostly occupied up to a certain level. That is, it is rather a system of many gravitational DoF being in a Dirac sea like state. However, the energy levels need not be negative as in the original Dirac sea.  This particular gravitational state which we designate by QST corresponds to a certain classical macro state, see the following section.
\begin{bem} This gravitational ground state is assumed to be a pure global quantum state, consisting of a complex network of local microscopic and more elementary states belonging to the individual gravitational DoF or small groups of them. We hence assume that it is possible to annihilate or alter some of these elementary micro states in this complex array, thus creating what may be viewed as a hole in the global state.
\end{bem}

\begin{assumption} If a QM excitation occurs sponaneously in the vacuum, a corresponding hole is assumed to occur in the QST partial state, i.e., the Dirac sea. This gravitational hole excitation corresponds to the negative mass excitation of the particle picture.
\end{assumption}
\begin{conclusion} By introducing this Dirac-sea concept of mostly occupied gravitational energy levels we avoid the necessary fine tuning of positive and negative matter-energy fluctuations in the vacuum as the spontaneous particle hole excitations do not alter the energy of the vacuum.
\end{conclusion} 

We thus see that this picture of an extended quantum vacuum state nicely fits with the zero energy universe concept we discussed above. Both pictures suggest the existence of a quantum mechanical coupling of QM excitations and quantum hole excitations in QST yielding a strongly correlated or entangled compound state. We will develop this picture in more detail in the next section.

%%%%%%%%%%%%%%%%%%%%%%%%%%%%
\section{Some Arguments in Favor of Quantum Hair and against the Correctness of the Semiclassical Picture}
In the preceding section we have seen that one should expect a close correlation between particle like excitations and corresponding gravitational hole excitations in QST.This already shows that we have to expect a notable interaction between QM and QST already on a microscopic level. On the other hand, the reason why in many of the approaches it is argued that local quantum field theory (LQFT) should hold (at least in the exterior region) is the \tit{nice slice} argument (for a more detailed discussion cf. for example \cite{Susskind1},\cite{Mathur1} or \cite{Giddings2}). Note however that these authors provide various arguments why this principle is presumably not entirely reliable. On the other hand, it is difficult to see what is exactly wrong with this seemingly harmless argument. Usually some kind of non-locality is invoked.
\begin{ob}If the nice slice argument is correct, Hawking's semiclassical analysis goes through and more intricate effects of quantum gravity can be ignored.
\end{ob}

 The standard argument is that the semiclassical approximation should work in regions where all aspects of curvature are sufficiently small and space-time (ST) is only slowly varying. This is in principle a low energy argument and winds up to:
\begin{assumption}If all the curvature and/or energy aspects of the scenario are sufficiently small, the coupling between matter and gravitational degrees of freedom (DoF) can be neglected or incorporated in a semiclassical manner.
\end{assumption}

In our view the criterium of low curvature or energy may be relevant for some quantitative and more macroscopic aspects of the problem but not! for the crucial question whether a pure ingoing state remains pure during the full evaporation process or goes over into a mixture. This is, quite to the contrary, not  a question of involved curvature energy but depends on other more microscopic  characteristics of the problem which have nothing to do with the largeness of the involved energies. What is actually of relevance is the way how matter and gravitational DoF entangle in the microscopic processes  which, as we will argue, is largely independent of the involved energy scale.

 In the BH case there exists, for one, almost by definition, a coupling between matter and gravitational DoF. A piece of matter which passes the event horizon falls within a finite lapse of time (either proper time or the time in an appropriate coordinate system like e.g. ingoing Eddington-Finkelstein, Novikov or Gullstrand-Painleve coordinates; cf. \cite{Gautereau} or \cite{Poisson}) into the singularity, thus changing the parameter $M$ in the Schwarzschild metric. By the same token the change in the metric should have an  effect on the corresonding quantum gravity microstate (see below). For an exterior observer beyond the event horizon this may happen even earlier, i.e. after the so-called \tit{scrambling time}, a concept which is currently under intense study (see e.g. \cite{Susskind2} or \cite{Hayden1}). This phenomenon of scrambling is by its very nature a desription of the averaged microscopic interaction of matter and gravitational DoF. The same holds for the creation of Hawking pairs out of the vacuum near the BH event horizon as discussed above. That is, there do exist several processes which suggest a kind of entanglement between matter and gravity DoF which may perhaps go beyond the semiclassical approximation.

%%%%%%%%%%%%%%%%%%%%%%%%%%%%
\section{(Q)ST as Order Parameter Manifold with g(x) as Order Parameter Field}
The preceding discussion suggests that we should treat the joint system of quantum matter and quantum space time on the microscopic level as a tensor product of Hilbert spaces.
\begin{propo}In the following we assume that our Hilbert space of matter plus gravity DoF can be represented as a tensor product, that is:
\beq \mcal{H}^{total}=\mcal{H}^m\otimes \mcal{H}^g  \eeq

\end{propo} 
 We analyze now in a next step the relation between macroscopic classical gravity and this underlying microscopic regime. In section 3 of \cite{Requ1} we described the classical metric tensor field
\beq g(x)=:<\tilde{g}(x)>          \eeq
as an \tit{orderparameter field}, being the expectation of a corresponding quantum observable, $\tilde{g}(x)$, of some underlying version of quantum gravity living in a certain Hilbert space. 
\begin{bem}We note that we make only very general assumption about this microscopic theory. We think every model of quantum gravity should fulfill these assumptions.
\end{bem} 
\begin{defi}We call such a classical $g(x)$ an order parameter field and the corresponding Lorentzian manifold $(g,\mcal{M})$ an {\em order parameter manifold} to emphasize the similarities to phase transitions and spontaneous symmetry breaking (SSB) in (quantum) many body physics.
\end{defi}
\begin{bem}These points are discussed in more detail in \cite{Requ2} and \cite{Requ1}, where we argued that gravitons are Goldstone modes of diffeomorphism invariance.
\end{bem} 

We now introduce another important concept which generalizes the concept of \tit{macro-observable}. This important idea was, as far as we know, for the first time developed by v.Neumann in a beautiful paper (\cite{Neumann1}), see also section V4, p212ff of the famous \cite{Neumann2}. As in statistical mechanics we associate to the macro-observable $g(x)$ a so-called \tit{phase cell}, denoted by $[g(x)]$, comprising all the quantum gravity micro-states, $\{\psi^g\}$, (of Hilbert norm one) with
\beq <\psi^g,\tilde{g}(x)\psi^g>=g(x)\quad\text{for all}\quad \psi^g\in\,[g(x)]    \eeq
\begin{bem}In statistical mechanics phase cells are frequently defined via the quantum many body states lying in a small energy interval $\Delta E$ between the energies $E$ and $E+\Delta E$ of some microcanonical distribution.
\end{bem}

Choosing now an orthonormal basis $\{\psi_k^g\}$ in the Hilbert subspace defined by $[g(x)]$ the general  state in $[g(x)]$ is given by
\beq \psi^g=\sum c_k\psi^g_k      \eeq
with arbitrary complex $c_k$ fulfilling $\sum |c_k|^2=1$.

One may treat in a first step the case where the (quantum) matter part is approximately independent of the (quantum) gravity part. That is, let $\phi^m$  be some state belonging to a quantum field theory model living in a curved ST background. As ordinarily the states of quantum field theory are assumed to be pure the states of matter and gravity cannot be entangled. This implies the following representation:
\begin{ob}In case matter and gravitational DoF are not entangled the corresponding joint quantum gravity state is a product state. i.e.:
\beq \Psi^{total}=\phi^m\otimes \psi^g    \eeq
with $\psi^g$ some state from $[g(x)]$. The partial trace over the gravity part is then again a pure state in the (quantum) matter Hilbert space.
\end{ob}
In the next section we will argue that this simple situation does not prevail in the presence of a BH. 
%%%%%%%%%%%%%%%%%%%%%%%%%%%%%
\section{The Entanglement of Matter and Gravitational DoF in the BH Scenario}
In the BH scenario we have to deal with quantum gravity in the exterior and the interior region behind the event horizon. The quantum theory in the interior region is notoriously delicate. We mention three representative papers which treat this problem (\cite{Susskind1},\cite{Giddings3},\cite{Mathur1}) for the case of a real non-stationary BH formed by some infalling matter. In \cite{Susskind1} nice slices are constructed (i.e. space-like Cauchy surfaces on which the ordinary LQFT program is supposed to work) by employing Kruskal-Szekeres coordinates. In the exterior region these slices are constant Schwarzschild time hypersurfaces which are then fitted smoothly in the interior region to the surface
\beq UV=R^2   \eeq
with $R$ chosen so large that the surface is nowhere near the singularity. The nice slices are parametrized by Schwarzschild time $t$ and one can introduce a weakly time dependent Hamiltonian which induces an evolution from one slice to another. In a next step a LQFT Hilbert space is constructed  in this low energy nice slice theory which decomposes as
\beq \mcal{H}(\text{slice})=\mcal{H}(\text{slice}_{int})\otimes\mcal{H}(\text{slice}_{ext})    \eeq
It is then argued that this approach fails with evidence for a certain non-locality coming from string theory.

In both \cite{Giddings3} and \cite{Mathur1} the exterior parts of the nice slices are as in \cite{Susskind1}. This part is then fitted in the interior to a $r=const$ part ($r$ the Schwarzschild $r$) which corresponds in the interior to constant coordinate time!). In both papers it is argued that the nice slice argument becomes unreliable because of the extreme stretching of the connector segment connecting the $r=const$ to the $t=const$ part for increasing time $t$ (cf. \cite{Mathur1}). Giddings in \cite{Giddings3} then introduces so-called \tit{natural slices} by choosing freely falling observers distributed in $r$ with initially synchronized clocks. This looks like using so-called \tit{Novikov coordinates}. The difference is now that the singularity is not avoided for all time as the innermost observer falls into the singularity in finite time. But as these slices are no longer Cauch surfaces we do not see how the ordinary quantization process can be performed in this construction. 
\begin{ob}In all these approaches it is argued that some version of non-locality slips in and that the nice slice construction does not give a fully adequate description of the quantum state of a BH. One should however note that in the above mentioned papers non-locality is not really derived from first priciples but is to some extent rather invoked to explain how information is able to leak out in the late radiation in order to prevent the final state to become a mixture.
\end{ob} 
\begin{bem}We noted in the introduction that in our view this problematical consequence of a certain non-locality results from using a pure particle picture. It can be avoided by complementing the radiation part discussed below by certain gravitational excitation modes. That is, the infalling particles denoted by the index c in the formula below are replaced by a suitable gravitational excitation mode which is located both inside and outside with respect to the event horizon.
\end{bem}  

In the above  mentioned papers the time dependence of the BH state and the evaporation process is entirely desribed via the formation of Hawking pairs, i.e. it is more or less a particle picture in a slowly varying classical background. In the toy model, discussed by Mathur (\cite{Mathur1}) the evolving state consists of the infalling matter state $\psi_M$ which is responsible for the creation of the BH and which is assumed to be sufficiently far away from the horizon regime and Hawking pairs:
\begin{multline}
 |\Psi>=|\psi_M>\otimes (2^{-1/2}|0>_{c_1}|0>_{b_1}+2^{-1/2}|1>_{c_1}|1>_{b_1})  \otimes \\ (2^{-1/2}|0>_{c_2}|0>_{b_2}+2^{-1/2}|1>_{c_2}|1>_{b_2})\cdots 
 \otimes \\ (2^{-1/2}|0>_{c_N}|0>_{b_N}+2^{-1/2}|1>_{c_N}|1>_{b_N})
\end{multline}
where the infalling particles are denoted by the index $c$ the particles gathering at infinity by $b$ (the Hawking pairs are assume to be generated one after the other).
It is then shown that such a picture leads to problems with conservation of information (i.e. unitarity). 
 We now want to proceed by arguing that this particle picture is not the whole story.

We proceed with the analysis which we started in the preceding section . With the classical metric tensor field as order parameter field and the Hilbert space of gravitational plus matter micro-states a tensor product
\beq  \mcal{H}^{total}=\mcal{H}^m\otimes \mcal{H}^g  \eeq
We split the tensor component $ \mcal{H}^g$ in the BH case further into:
\beq \mcal{H}^g=\mcal{H}^g_{int}\otimes\mcal{H}^g_{ext}    \eeq
but this turns out to be a subtle point. This means the following. In various papers such a picture is invoked because on a Cauchy hypersurface in LQFT spacelike operators and the res pective subalgebras do commute and it is sometimes argued that this is already sufficient for having such a product structure on the level of states. But this is not correct in general. In ordinary LQFT in Minkowski space it is known that only a weaker result does hold (see e.g. sect. V.5.2 in \cite{Haag} or the detailed discussion in \cite{Requ4}. 

In the case of the Unruh effect it is known that the wide spread representation of the Minkowski vacuum as a vector in the tensor product of  left and right wedge tensor factors is, strictly speaking, incorrect. This can easily be seen because one has in this scenario the Minkowski vacuum state as a reference state. On the other hand, we know that the generator of Lorentz boosts in the wedges is the corresponding KMS Hamiltonian, having continuous spectrum, while in the tensor product the logarithm of the density matrix is the canonical Hamiltonian thus having a discrete spectrum (cf. \cite{Requ4} and sect. 3 of \cite{Requ5}). In the BH scenario such a standard reference state is missing and one may conjecture that the fact that the BH event horizon (Schwarzschild case) is a \tit{Killing horizon}, i.e. a \tit{null surface} to which a Killing field is normal, justifies such an assumption.
\begin{bem}Our conclusions in the following do however not really depend on this perhaps too restrictive assumption.
\end{bem}
%%%%%%%%%%%%%%%%%%%%%%%%%%%%%%
\subsection{The Pure Gravity Case}
In a first step we treat only pure (quantum) gravity states and for reasons to be explained immediately we make the following assumption:
\beq \mcal{H}^g\subset\mcal{H}^g_{int}\otimes\mcal{H}^g_{ext}    \eeq
While the full (mathematical) basis in the tensor product is 
\beq \{\psi^{int}_k\otimes\psi^{ext}_l\}\quad k,l=1,\cdots ,N     \eeq
the admissible quantum gravity states span only the following subspace of $\mcal{H}^g_{int}\otimes\mcal{H}^g_{ext}$:
\beq \mcal{H}^g= span\{\psi^{int}_k\otimes\psi^{ext}_k\}     \eeq
\begin{bem}Remember that in the Rindler case the Minkowski vacuum $\Omega$ is such a strongly entangled state
\beq \Omega=\sum c_k\psi^L_k\otimes\psi^R_k    \eeq
\end{bem}
We make now the following natural assumption. We assume that there exists some microscopic unitary ``time'' evolution, $U_{\tau}$, in $\mcal{H}^g$, an evolution we as macroscopic observers are presumably unable to observe. 
\begin{bem}We note that time is a notoriously difficult concept in (quantum) gravity. One should regard this evolution parameter rather in the spirit as formulated in e.g. \cite{Susskind1}.
\end{bem}
\begin{assumption}Under this evolution the $\psi^{int}_k\otimes\psi^{ext}_k$ are assumed to be eigenstates, i.e., 
\beq U_{\tau}( \psi^{int}_k\otimes\psi^{ext}_k)=e^{iE_k\tau}(\psi^{int}_k\otimes\psi^{ext}_k)  \eeq
\end{assumption}

The above subspace property is a very important point which needs some clarifying remarks.
In our view the aspects of a certain non-locality, which were alluded to in the above cited papers and which were partly attributed to string theory, have their roots in the deep structure of the quantum gravity vacuum. The \tit{Bekenstein-Hawking BH entropy area law} tells us that the number of gravitational microscopic DoF is proportional to the area of the event horizon. Furthermore, irrespectively of the possibility that string theory may be the \tit{only game in town}, we developed in \cite{Requ6} a \tit{translocal} structure in the quantum vacuum called by us \tit{wormhole spaces} and which leads to such an area law whereas the respective DoF are lying both in the interior/exterior and on the dividing horizon. That is, we conclude the following:
\begin{conclusion}Both $\mcal{H}^g_{int}$ and $ \mcal{H}^g_{ext}$ have the same dimension (in natural units)
\beq N=A/4    \eeq
with $A$ the area of the event horizon.  If the microstate in the interior is some $\psi^{int}_k$ it is strictly coupled to (or entangled with) via the corresponding boundary conditions on the event horizon to the respective mirror state $\psi^{ext}_k$, i.e. the total basis is spanned by vectors $\psi^{int}_k\otimes\psi^{ext}_k$ and a pure quantum gravity state in the BH case reads
\beq \label{BH} \psi^g=\sum c_k\psi^{int}_k\otimes\psi^{ext}_k      \eeq
\end{conclusion}
\begin{bem}Note that this dimension $N$ shrinks in the evaporation process; for the details see below. This implies of course that the respective basis vectors also change. This means that in the following we are explicitly using a representation of states which depend on some evolution parameter. That is, we find it difficult in this context to employ the standard Heisenberg picture. The latter framework might perhaps work in some super-Hilbertspace formed by all the elementary quantum gravity DoF and in which all the evolution parameter dependent Hilbertspaces we are using are contained but, in any case, this would be a quite cumbersome picture in our view. (as to the use of the Heisenberg picture in curved space-time see e.g. the nice reviews \cite{Ford} and \cite{Isham}).
\end{bem}

From this observation already follow some important corollaries:
\begin{koro}~\\
i) Due to our wormhole picture an interior bulk state $\psi^{int}_k$ leads to a unique boundary condition on or in the vicinity of the event horizon. This boundary condition is standing in a one-one relation to the corresponding bulk states $\psi^{int}_k$ and $\psi^{ext}_k$. This relation expresses on the one hand the peculiar long-range correlation or entanglement between interior and exterior states. On the other hand, it shows that the relevant gravitational microscopic DoF are located as well in the interior, exterior and near the event horizon of the BH.\\
ii)Thus the microscopic boundary state fixes both the corresponding interior and exterior bulk state and is thus a realization of the {\em holographic principle}.
\end{koro}
These observations clarify in our view some of the long standing questions raised in \cite{Bekenstein1},\cite{Bekenstein2},\cite{Wald},\cite{Jacobson},\cite{Bousso}. I.e., while the BH microstates are of an extended bulk nature in our picture, they behave, due to the \tit{holographic} strong \tit{bulk-boundary} coupling also as \tit{surface states}, that is, the bulk states can be labelled by the respective boundary conditions.

There is another problem which was widely discussed in the past, namely the microscopic nature of the BH entropy. That is, is it pure entanglement entropy, is entanglement entropy only a part of it or is BH entropy something entirely different. This problem was raised for example by the results in  \cite{Bombelli} and \cite{Srednicki} and discussed in \cite{Bekenstein1} and \cite{Bekenstein2}. From the general representation of a BH microstate in the form of formula 
(\ref{BH}) we can infer the following. As the strong entanglement implies that the BH microstate is automatically in the Schmidt form, the calculation of the entanglement entropy is simple. By tracing over the exterior or interior components in the state one gets
\beq S_{ent}=-\sum |c_k|^2\ln |c_k|^2   \eeq
\begin{bem}Note that in the Schmidt representation we can arrange matters so that the $c_k$ are real and positive by absorbing phase factors in the basis vectors.
\end{bem}
The above expression for the entropy is what we call in statistical mechanics \tit{microscopic entropy} which is constant under the hypothetical evolution $U_{\tau}$. It is common practice to work in such a context rather with some averaged or coarse grained entropy as the microscopic entropy is presumably unobservable anyhow.

It was the explicit philosophy of the ingeneous paper \cite{Neumann1} to establish such a program. The coarse graining was accomplished by defining certain \tit{macro-observables} which yield a division of quantum mechanical phase space into \tit{phase cells}. In our context such a macro observable is given by the classical metric tensor
\beq g(x)=<\tilde{g}>    \eeq
which defines the corresponding Hilbert subspace 
\beq  \mcal{H}^g=[g(x)]    \eeq
as phase cell. Thus, in the spirit of \cite{Neumann1} the averaging should be performed over $\mcal{H}^g$. 

A \tit{random vector} in  $\mcal{H}^g$ is given by
\beq \psi^g=\sum c_k\psi^{int}_k\otimes\psi^{ext}_k   \eeq
with the $c_k$ varying over the unit hypersphere $S^{2N-1}$ ($N$ the complex dimension of $\mcal{H}^g$).  That is, the average is performed over $S^{2N-1}$ with the normalized Haar measure of the respective rotation group. We begin the discussion with a useful lemma:
\begin{lemma}With $Exp_{S^{2N-1}}$ the integration over $S^{2N-1}$ with respect to the normalized Haar measure on $S^{2N-1}$ and $\rho_{mc}$ the density matrix of the microcanonical distribution function over $\mcal{H}^g$ it holds:
\beq   Exp_{S^{2N-1}}(<\psi^g,A\psi^g>)=tr(\rho_{mc}A)    \eeq
\end{lemma}
Proof: The proof involves the evaluation of various integrals over $S^{2N-1}$ of the type
\beq \int \overline{c}_i\cdot c_j\,d\mu=\delta_{ij}\cdot\int |c_i|^2\,d\mu=1/(2N-1)    \eeq
using symmetry arguments (cf. Appendix II in the nice book \cite{Jancel}).

We remark that there does exist a very active field of current research in which such types of arguments are used, having great relevance for various fields of modern physics, catchwords being \tit{typicality} and \tit{concentration of measure phenomenon}. To our knowledge such research was initiated by von Neumann's paper \cite{Neumann1}. In statistical mechanics see for example \cite{Lebowitz1},\cite{Lebowitz2},\cite{Popescu}. Having some relevance to BH physics we cite the fundamental \cite{Page2}, see also \cite{Hayden1}, a recent more mathematically oriented book is \cite{Ledoux}.
The above result now motivates the average entropy of a random quantum gravity microstate to be defined by the corresponding microcanonical entropy of the Hilbert subspace $\mcal{H}^g$, i.e.:
\begin{ob}It follows from our above reasoning for the entropy of a random micro-state in $\mcal{H}^g$ that
\beq S(\psi^g_{random})=N    \eeq
that is, it is the dimension of the subspace $\mcal{H}^g=[g(x)]$ and which we associate with the Bekenstein-Hawking BH entropy.
\end{ob}
On the other hand, this is the entanglement entropy of the maximally entangled microstate:
\beq \psi^g=\sum N^{-1/2}\psi^{int}_k\otimes\psi^{ext}_k    \eeq
\begin{bem}Note that this is the kind of entropy which one should attribute to a state with given $g(x)=<\tilde{g}(x)>$ and assuming maximal ignorance about the further details of the microstate according to the philosophy of Jaynes (\cite{Jaynes}).
\end{bem} 
%%%%%%%%%%%%%%%%%%%
\subsection{The Complete Quantum Gravity plus Radiation State}
We now come to the crucial part of this paper. For technical convenience we make the simplifying assumption that our BH starts from a pure quantum gravity initial state 
\beq \psi^g(0)=\sum_k c_k(0)\psi^{int}_k(0)\otimes\psi^{ext}_k(0)    \eeq
We assume (as explained above) that we have some evolution parameter at our disposal. See for example \cite{Susskind1} for an evolution which labels the sequence of consecutive nice slices. We furthermore assume that Hawking particles are created near the event horizon with some particles escaping to infinity, some others falling into the BH. As to these particles falling into the BH and reaching the singularity  in finite proper time we shall be a little bit cavalier  in the following. It depends on the type of coordinate system and slices we are using how to describe properly their detailed evolution in the interior of the BH. In e.g. \cite{Mathur1} they are accumulating, according to the particular construction of the nice slices, on the part of the slices defined by $r=const< 2M$ ($r$ the Schwarzschild radial coordinate). In other coordinates or scenarios they fall into the singularity, but note the following:
\begin{bem}Unfortunately the concept of particle is heavily dependent on the choice of coordinate system near the event horizon (see the beautiful analysis in e.g. \cite{Unruh2} or the remarks in \cite{Wald2}. This becomes particularly apparent when applying the \tit{equivalence principle}, i.e. using infalling observers or infalling coordinates and comparing infalling observers with observers at constant Schwarzschild coordinate $r$.
\end{bem}    

Therefore we make, for convenience, the following choice in the sequel. We treat $\psi^{int}_k$ as a compound state, describing both the quantum gravity microstate in the interior plus possible particle excitation modes being present behind the event horizon at evolution time $\tau$. As our main focus lies on what happens outside the event horizon, this simplification will not be of much concern.  

In contrast to the semiclassical analysis we shall now explicitly take into account the structural change in the quantum gravity microstates due to the presence of the created Hawking particles moving either to infinity or towards the singularity, that is, the so-called \tit{backreaction}. Furthermore there is the effect on the parameters of the BH by the particles which vanished in the singularity (reducing the mass $M$). That is, at evolution ``time'' $\tau$ we assume the following:
\begin{propo}At evolution ``time'' $\tau$ we have the following compound state
\beq \psi^{g,rad}=\sum_i(\sum_k c_{ki}(\tau)\psi_{ki}^{int}(\tau)\otimes\psi_{ki}^{ext}(\tau))\otimes\phi^{rad}_i(\tau)    \eeq   
where both the individual tensor components, the coefficients and the range of summations (i.e. the number of terms) depend on the evolution parameter $\tau$. The $\psi_{ki}^{int}(\tau)\otimes\psi_{ki}^{ext}(\tau)$ represent quantum gravity microstates in the Hilbert subspace being labelled by the exterior particle state $\phi^{rad}_i(\tau)$. It is however important that they no longer are members of an ON-basis. They are assumed to be perturbations of the corresponding BH microstates $\psi_{k}^{int}(\tau)\otimes\psi_{k}^{ext}(\tau)$, that is, the microstates of a BH with the macroscopic parameters which prevail at evolution ``time'' $\tau$ but without the radiation states.
\end{propo}
\begin{bem}Note that this entanglement between particle and gravitational quantum states comes about due to the interaction mechanism between QM-excitations and gravitational QST-excitations we described in the first sections of our paper.
\end{bem}

One can model this type of interaction if one assumes that some form of interaction Hamiltonian $H_I$ does exist.
\begin{propo}With $a_i^{\dagger}$ creating some QM-excitation modes and $b_i$ annihilating the corresponding gravitational excitation modes in QST we assume some interaction Hamiltonian of the form
\beq  H_I=\lambda\sum_i (a_i^{\dagger}b_i+a_ib_i^{\dagger})   \eeq
\end{propo}

Some clarifying remarks are in order. The particles entering the interior of the BH change of course the microscopic gravitational field there as do the particles vanishing in the singularity and, by the same token, the macroscopic, i.e., classical $g$-field. The same holds for the exterior field. As the interior becomes smaller in the course of evolution and evaporation and, correspondingly, the exterior larger, the dimension of the Hilbert subspace $\mcal{H}^g(\tau)$ which is spanned by the vectors $\psi_{k}^{int}(\tau)\otimes\psi_{k}^{ext}(\tau)$ will shrink. For the final state we have:
\begin{koro}This implies that the  $\psi_{ki}^{int}(\tau)\otimes\psi_{ki}^{ext}(\tau)$ (the perturbations of the  $\psi_{k}^{int}(\tau)\otimes\psi_{k}^{ext}(\tau)$) for an almost evaporated BH span a subspace of a very small dimension whereas the summation over the index $i$ may be quite large. 
\end{koro}
\begin{ob}When the BH has completely evaporated, the final pure! state which remains is
\beq  \psi^{g,rad}_{final}=\sum_i\, c_{f,i}\psi_{f,i}\otimes\phi_i^{rad}    \eeq
with the $\psi_{f,i}$  being perturbations of the single quantum gravity state $\psi_f$. We can now again assume that the $\psi_{f,i}$ are members of an ON-basis, so the  partial trace yields
\beq \rho_f=|c_i|^2|\phi_i^{rad}><\phi_i^{rad}|     \eeq
This partial trace is assumed to coincide with the thermal final state of the Hawking semiclassical analysis.
\end{ob} 
\begin{bem}The effect of the existence of a BH is hence that the final pure state is not a product state as in simple situations but a superposition of such tensor product states.
\end{bem}  
%%%%%%%%%%%%%%%%%%%%%
\section{Conclusion}
We have argued that the semiclassical analysis of the BH information paradox is incomplete as are ultimately all the variants which are based on it. This incompleteness derives from the neglection of the quantum gravity DoF which are entangled with the matter DoF irrespective of the fact that the involved energies may be low. We follow the pure entangled matter-gravity compound state through its full evolution until the complete evaporation of the BH, arriving at a final pure state which, if traced over the quantum gravity tensor component, yields the thermal Hawking radiation pattern. Our analysis of the nature of the entanglement of matter and quantum gravity DoF is also relevant in a wider context of similar scenarios.  It should be emphasized that our analysis does not rely on a particular model theory like e.g. string theory.

\end{document}